\RequirePackage{fix-cm}
\documentclass[smallextended]{svjour3}       
\smartqed  
\usepackage{graphicx}
\usepackage{pst-plot}
\journalname{Mathematical Programming}
\usepackage{amsmath}
\usepackage{amssymb}
\usepackage[ruled]{algorithm} 
\usepackage{srcltx}
\usepackage{epsfig}

\def\RR{{\mathbb R}}

\def\EE{{\mathbb E}}
\def\PP{{\mathbb P}}
\def\vphi{{\varphi}}
\def\p{\prime}

\def\la{\lambda}
\def\eps{\epsilon}
\def\ind{e}
\def\EP{\mathcal{E}}

\newtheorem{prop}{Proposition}
\newtheorem{lema}{Lemma}
\newtheorem{teo}{Theorem}
\newtheorem{defi}{Definition}

\begin{document}

\title{Network Congestion Control with Markovian Multipath Routing
\thanks{Partially supported by FONDECYT 1100046 and Instituto 
Milenio Sistemas Complejos de Ingenier\'ia.}
\footnote{A preliminary short version of this paper was published
in the Proceedings of the 5th International Conference on Network Games, 
Control and Optimization (NetGCOOP'2011, Paris).}
}


\author{Roberto Cominetti         \and
        Crist\'obal Guzm\'an 
}


\institute{R. Cominetti \at
              Departamento de Ingenier\'ia Industrial, \\
              Universidad de Chile, Rep\'ublica 701,\\
              Santiago, 8370439 Chile\\
              \email{rccc@dii.uchile.cl}           
           \and
           C. Guzm\'an \at
		School of Industrial and Systems Engineering\\
		Georgia Institute of Technology, 765 Ferst Drive, NW,\\
		Atlanta, GA 30332--0250, USA\\	
		\email{cguzman@gatech.edu}
 }

\date{Received: date / Accepted: date}

\maketitle

\begin{abstract}
In this paper we consider an integrated model for TCP/IP protocols with multipath routing.
The model combines a Network Utility Maximization for rate control based on end-to-end queueing 
delays, with a Markovian Traffic Equilibrium for routing based on total expected delays. We prove 
the existence of a unique equilibrium state which is characterized as the solution of an unconstrained 
strictly convex program. A distributed algorithm for solving this optimization problem 
is proposed, with a brief discussion of how it could be implemented by adapting the current Internet protocols. 
\end{abstract}

\section{Introduction}
\label{intro}

Routing and congestion control are two basic components of packet-switched communication
networks. While {\em routing} is responsible for determining efficient paths along 
which the sources communicate to their corresponding receivers, {\em congestion control}
manages the transmission rate of each source in order to keep network congestion within reasonable limits. 
In current practice both mechanisms belong to separate design layers that operate on different 
time-scales: the IP layer (Internet Protocol) determines single-path routings which are updated on a 
slow time-scale, while the TCP layer (Transmission Control Protocol) corresponds to end-to-end users 
that perform rate congestion control at a faster pace for which routing can be considered to be fixed.  
Scalability  considerations impose that these protocols must operate in a decentralized manner.

Roughly speaking, TCP controls the rate of a source by managing a {\em window size} that bounds the maximum 
number of outstanding packets that have been transmitted but not yet acknowledged by the receiver. 
Once this window size is reached the source must wait for an acknowledgment before sending a new 
packet, so the rate is approximately one window of packets per round-trip time. As the network 
gets congested, the round-trip time increases and the transmission rate is automatically slowed down. 
In addition, TCP dynamically adjusts the window size of a source in response to network congestion. 
To this end, links generate a scalar measure of their own congestion (e.g.
packet loss probability, average queue length, queueing delay) and each source is fed back a
{\em congestion signal} that reflects the aggregate congestion of the links along its route.
This signal is used by the source to adjust its window size so that the higher the congestion the 
smaller the rate. The predominant TCP protocols in use are Tahoe and Reno which use 
{\em packet loss} as congestion measure, and Vegas which is based on {\em queueing delay}. 
We refer to \cite{Low:2002} for a description and comparison of current protocols and their models.

The interaction of many sources performing a decentralized congestion control based 
on feedback signals that are subject to estimation errors and communication delays, gives rise to complex dynamics 
that are difficult to analyze. 
However, assuming that the dynamics stabilize on a steady state, the equilibrium can be characterized as an 
optimal solution of a Network Utility Maximization (NUM) problem (see \cite{Kelly:1998,Low:2003,YMR98}).
Thus, the TCP mechanism can be viewed as a decentralized algorithm that seeks to 
optimize an aggregate utility function subject to network constraints. The NUM approach 
is also useful to compare different protocols in regard to their fairness and efficiency. 

A second element of packet-switched networks is {\em routing}. This function is performed by routers in a 
decentralized manner using routing tables that determine the next hop for each destination. The routing tables 
are updated periodically by an asynchronous distributed shortest path iteration that computes optimal paths 
according to some metric such as hop count, latency, delay, load, reliability, bandwidth, or a mixture of these. 
In current practice a single path that minimizes the number of hops is used for routing on each origin-destination pair.
A very promising idea in traffic engineering is the use of multipath routing to increase the throughput by exploiting 
the available transmission capacity on a set of alternative paths. This also improves the reliability because 
of the ability to redirect flow into alternate paths  in case of failures.
While a few multipath techniques are available in today's Internet ({\em e.g.} 
MPLS tunnels \cite{Villamizar:1999} or MPTCP multihoming developed by IETF \cite{Barre:2011}), 
several other proposals have been tested through simulations. We defer our discussion of the relevant literature 
until section \S\ref{related-work}.
Another relevant argument in favor of multipath routing is stability. 
Indeed, when route choice is based on metrics that are affected by congestion, such as queueing delay or link latencies, routing and rate control become mutually 
inter-dependent and equilibrium can be achieved only if both aspects are considered jointly: routing 
affects the rate control through the induced congestion signals, while  rate control 
induces flows that determine in turn which routes are optimal. If routing is restricted to a single path, 
congestion effects may lead to route flaps. A remedy for such unstable behavior is to allow flows to split 
over multiple paths in order to balance their loads. An appropriate tool to capture these interactions 
between rate control and routing is provided by Wardrop equilibrium.  
On the other hand, since congestion metrics are subject to estimation errors and random effects it is
natural to model routing as a stochastic equilibrium assignment.

The goal of this paper is to propose an analytical framework that
provides a theoretical support for cross-layer designs for rate 
control under multipath routing. Our model combines rate control 
modeled by NUM with a routing strategy based on discrete choice distribution models 
that lead to a Markovian Traffic Equilibrium (MTE).  The latter is a decentralized stochastic 
version of Wardrop's model.  
The combination of the NUM and MTE models leads to a system of equations that correspond to the 
optimality conditions of an equivalent Markovian Network Utility Maximization problem (MNUM),
a strictly convex unconstrained program of low dimension where the variables are the 
link congestion prices. This characterization allows us to establish the existence and uniqueness 
of an equilibrium, and provides a basis for designing decentralized protocols for congestion 
control and multipath routing.

The paper is structured as follows. Section \S\ref{prelim}  reviews the basic components
of our cross-layer approach: we recall the NUM framework for modeling  
the steady state of TCP protocols and we discuss the concepts of Wardrop equilibrium and Markovian routing. 
Section \S\ref{cross-layer} combines NUM and MTE, introducing the MNUM model for routing and rate control.
In \S\ref{reduced-model} we reduce MNUM to a system of equations involving only the link 
congestion prices, and then in \S\ref{variational-model} we show that these equations admit a variational
characterization {\sc (d-mnum)} proving the existence of a unique equilibrium state. 
In \S\ref{distributed-algorithm} we briefly discuss how the model might lead 
to a cross-layer design of a distributed TCP/IP protocol. We close the paper with comparisons
to previous work  and some perspectives on future research.

\section{Notations and preliminaries}
\label{prelim}

The communication network is modeled by a directed graph $G=(N,A)$, where the nodes 
$i\in N$ represent origins, destinations and intermediate routers, while the arcs $a\in A$ 
represent the network links.  
Each link is characterized by a    latency function $\la_a=s_a(w_a)=\lambda_a^0+\rho_a(w_a)$
where $\lambda_a^0\geq 0$ represents a constant propagation delay and $\rho_a(w_a)$ 
is the expected queueing delay expressed as a  continuous and strictly increasing 
function  $\rho_a:[0,c_a)\to[0,\infty)$ of the traffic $w_a$ on the link,  
with $\rho_a(0)=0$ and $c_a\in (0,\infty]$ the maximal capacity.
We also consider a finite set of sources $k\in K$ each one generating
a flow rate $x^k\geq 0$ from an origin $s_k\in N$ to a destination $d_k\in N$.

\subsection{Rate control and utility maximization under single path routing}
\label{NUM}
Suppose that each source $k\in K$ routes its flow along a fixed sequence of links $(a_1,\ldots,a_{j_k})$,
so that  the total traffic on a link $a$ is $w_a=\sum_{k\ni a}x^k$ where the 
summation is over all the sources $k\in K$ whose route contains that link. 
Consider the queueing delay 
$p_a=\rho_a(w_a)$ as a measure of link congestion and
assume that each source $k\in K$ adjusts its rate $x^k=f_k(q^k)$ as a function of 
the aggregate queueing $q^k=\sum_{a\in k}p_a$ on its route, 
 where $f_k:(0,\infty)\to(0,\infty)$ is continuous and strictly decreasing
with $f_k(q^k)\to 0$ as $q^k\to\infty$. 
These equilibrium equations may be written as
$$f_k^{-1}(x^k)=q^k=\sum_{a\in k}p_a=\sum_{a\in k}\rho_a(w_a)=\sum_{a\in k}\rho_a(\mbox{$\sum_{s\ni a}x^s$})$$
which correspond to the optimality conditions for the strictly convex program
$$
\min_{x\in\RR^K}  \sum_{a\in A} R_a(\mbox{$\sum_{s\ni a} x^s$})-\sum_{k\in K} \int_0^{x^k}\!\!\!\!f_k^{-1}(z)dz
\leqno{\mbox{\sc (num)}}
$$
where $R_a(\cdot)$ denotes a primitive of  $\rho_a(\cdot)$. Alternatively, the equations may be stated in terms of the 
queueing delays as
$$\rho_a^{-1}(p_a)=w_a=\sum_{k\ni a}x^k=\sum_{k\ni a}f_k(q^k)=\sum_{k\ni a}f_k(\mbox{$\sum_{b\in k}p_b$})$$
which are the optimality conditions for the strictly convex dual program
$$\min\limits_{p\in\RR^A}\sum\limits_{a\in A} \int_0^{p_a}\!\!\!\rho_a^{-1}(y)\,dy - \sum\limits_{k\in K} F_k(\mbox{$\sum_{b\in k}p_b$})\leqno{\mbox{\sc (d-num)}}$$
where $F_k(\cdot)$ is a primitive of $f_k(\cdot)$.

\vspace{2ex}
\noindent{\sc Example.}
Consider the model for TCP Vegas proposed in \cite{Low:2003,LPW:2002}. For each source $k$ 
and time $t$, let $W^k_t$ denote the size of the congestion window and
$T^k_t=D^k+q^k_t$ the RTT expressed as the sum of the total propagation delay 
$D^k$ and the queueing delay $q^k_t$.
A Vegas source estimates $D^k$ as the minimum observed RTT, and tries to keep the difference 
between the {\em expected rate} $\hat x^k_t=W^k_t/D^k$ and the {\em actual rate} $x^k_t=W^k_t/T^k_t$ 
close to a given value $\alpha^k>0$. To this end, the congestion window  is increased  if 
$\hat x^k_t-x^k_t<\alpha^k$, and decreased when $\hat x^k_t-x^k_t>\alpha^k$. At equilibrium we must have
$\hat x^k-x^k=\alpha^k$ which yields the equilibrium rate functions
$$x^k=\frac{\alpha^kD^k}{q^k}\triangleq f_k(q^k). $$
A simple model for queueing delay can be obtained by considering each link  as an M/M/1  queue
with service rate $c_a>0$ and an infinite buffer, which gives the expected
queueing delay 
$$p_a=\frac{w_a}{c_a(c_a-w_a)}\triangleq\rho_a(w_a).$$
The {\sc (num)} formalism can handle other congestion measures besides queueing delay 
and has been used to model the 
steady state of different TCP protocols, each one characterized by specific maps  $f_k$ and $\rho_a$
(see \cite{Dumas:2001,Gibbens:1999,Kelly:1998,Kunniyur:2003,Low:2002,Padhye:2000}).

\subsection{Routing and traffic equilibrium}\label{s22}
We review next some equilibrium models for traffic routing in congested networks. In this setting the source flow 
rates $x^k$ are fixed but may be routed along a set of alternative paths $R^k$ connecting the origin $s_k$ to the 
destination $d_k$. The basic modeling principle introduced by Wardrop in \cite{Wardrop:1952} is that at equilibrium 
only paths that are optimal should be used to route flow.

In contrast with rate control which uses queueing delay $p_a=\rho_a(w_a)$, 
route optimality will be measured using the total delays $\la_a=\la_a^0+\rho_a(w_a)$
so that packets are routed along paths with smaller round trip times and not only
small queuing delays. The rationale is that the earlier each packet is delivered, 
the larger the rate. The protocol should automatically select the most efficient routes, depending on 
the congestion prevailing on each link. A further advantage of choosing the currently shortest path 
using total delay is to ensure that packets arrive in order to their destination, reducing 
the conflicts with the duplicate {\em ack} mechanism  for detecting packet losses in TCP.

\vspace{2ex}
\noindent 
{\sc Example.}
To illustrate the point, consider two parallel links with identical queuing capacity but one of them 
with much longer propagation delay. A routing based on queuing delay alone would yield a 50\% 
traffic split, with {\em ack} delays dominated by  the slow link which unnecessarily limits the 
transmission rate (the fast link being under-utilized). Instead, a routing based on total delay will 
use the fast link more intensively until increased queuing makes the slow link competitive, 
achieving a higher throughput. Naturally, the fast link will have a larger queue as compared to 
the slow link, but not larger than in single-path routing as long as TCP is still controlling the amount 
of traffic using the Vegas mechanism. Eventually, a very slow link will not be used at all, which is 
again consistent with supporting higher rates.

\subsubsection{Wardrop equilibrium}
Suppose that the flow $x^k$ is split into non-negative path-flows $h_r\geq 0$ so that
$x^k = \sum_{r\in R^k} h_r$, and let $w_a= \sum_{r\ni a} h_r$ be the induced total link-flows.
Let $H$ denote the set of such {\em feasible flows} $(h,w)$.
An equilibrium \cite{Wardrop:1952} is characterized by the fact that only 
optimal paths are used, namely, for each destination $k\in K$ and each route $r\in R^k$ one has
\begin{equation}\label{WE}
h_r>0 \Rightarrow c_r=\tau_k
\end{equation}
where $c_r=\sum_{a\in r} \la_a=\sum_{a\in r} s_a(w_a)$ denotes the total delay of the route
and $\tau_k=\min_{r\in R^k} c_r$ is the minimum cost faced by source $k$. 

These equilibria were characterized in \cite{Beckmann:1956}  as the optimal solutions
of the convex program
$$
\min\limits_{(h,w)\in H} \sum\limits_{a\in A} \int_0^{w_a} \!\!\!s_a(z)\,dz.\leqno{\mbox{\sc (p-w)}}
$$
Since the feasible set $H$ is compact this problem has optimal solutions, 
while strict convexity implies that the optimal $w$ is unique. Alternatively, the 
equilibrium delays $\la_a=s_a(w_a)$ are the unique optimal solution 
of the strictly convex unconstrained  dual problem
$$
\min\limits_{\lambda\in\RR^A} \sum\limits_{a\in A} \int_{\la_a^0}^{\la_a} \!\!\!s_a^{-1}(z)\, dz-\sum_{k\in K}x^k  \tau_k(\la)\leqno{\mbox{\sc (d-w)}}
$$
where $\la_a^0=s_a(0)$ and $\tau_k(\la)\triangleq\min_{r\in R^k}\sum_{a\in r}\la_a$ is the minimum
total delay for source $k\in K$.

\vspace{2ex}
\noindent{\sc Remark.} The well known Braess' Paradox indicates that there are situations in which
forbidding flow on some links might lead to a modified equilibrium where all sources benefit from 
smaller travel times. This raises the relevant question of which links should be forbidden to 
optimize network performance. Unfortunately, as shown in \cite{Roughgarden:2006} 
this problem turns out to be hard to solve
even approximately and even for a single source.
Namely, unless P=NP, there is no polynomial approximation algorithm with approximation ratio less 
than $n/2$ where $n$ the number of nodes, while the optimal ratio $n/2$ is trivially attained
by forbidding no link.  As a consequence, {\em harmful} links cannot be
detected efficiently. 
To compensate, it is worth mentioning that for sufficiently 
high levels of demand and congestion, Braess' Paradox does not occur (see \cite{Nagourney:2010}).

\subsubsection{Markovian routing and equilibrium}\label{Markov_routing}

When link delays are subject to stochastic variability, the route delays  $\tilde c_r$ 
become random variables and the equilibrium conditions \eqref{WE} 
are replaced by a stochastic assignment of the form
 $h_r=x^k\,\PP(\mbox{$\tilde c_r$ is optimal})$. 
For instance, if the costs $\tilde c_r$ are i.i.d. Gumbel variables with expected value
$c_r=\EE(\tilde c_r)$, we get the Logit distribution rule common in the transportation literature
$$h_r=x^k{\exp(-\beta c_r)\over \sum_{p\in R^k}\exp(-\beta c_p)} \quad \quad(\forall\,r\in R^k)$$
which assigns flow to all the paths, favoring those with smaller expected cost $c_r$. 
The parameter $\beta$ controls how concentrated is the
repartition: for $\beta\sim 0$ every path receives an approximately equal share of the flow, while 
for $\beta$ large the flow concentrates on paths with minimal cost.
 Unfortunately, given the exponential number of end-to-end paths, 
such route-based distribution rules controlled directly by sources do not seem
amenable to design decentralized scalable routing protocols. This becomes 
critical if the protocol is expected to be responsive when facing route delays that vary with
traffic congestion.

\begin{figure}[!t]
\label{flow_cons}
\centering
\includegraphics[scale=0.4]{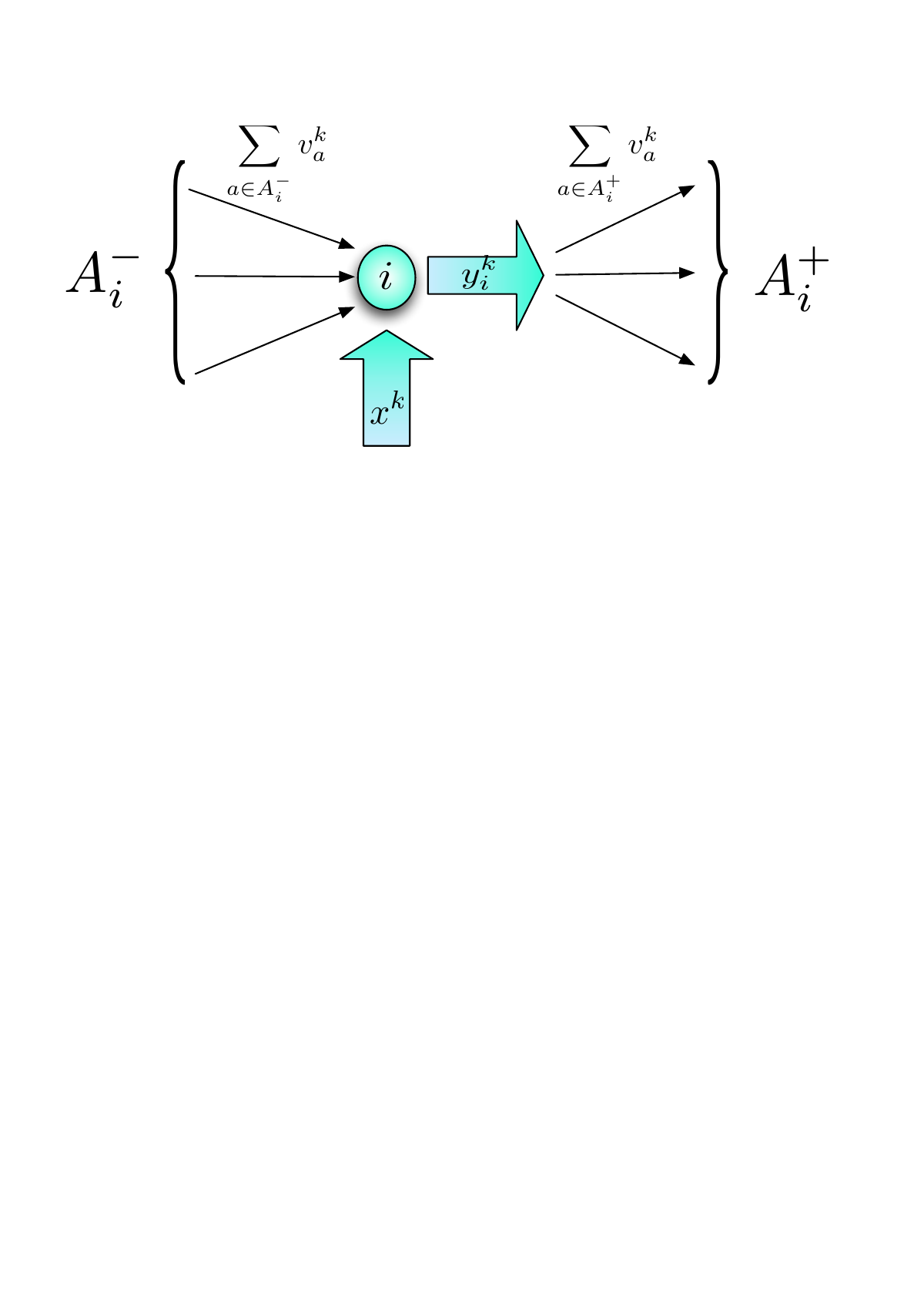}
\vspace{-2ex}
\caption{\label{dp2}Flow conservation diagram (here $i=s_k$)}
\end{figure}

An alternative is to conceive routing as a stochastic dynamic programming process. Suppose that
each packet experiences a random delay $\tilde \la_a$ when traversing link $a$, and 
let  $\tilde\tau_i^k$ be a random variable that represents the total delay from node $i$ to 
destination $d_k$. Denote $\la_a=\EE(\tilde\la_a)$ and $\tau_i^k=\EE(\tilde\tau_i^k)$ their 
expected values. 
\begin{figure}[!t]
\centering
\includegraphics[scale=0.38]{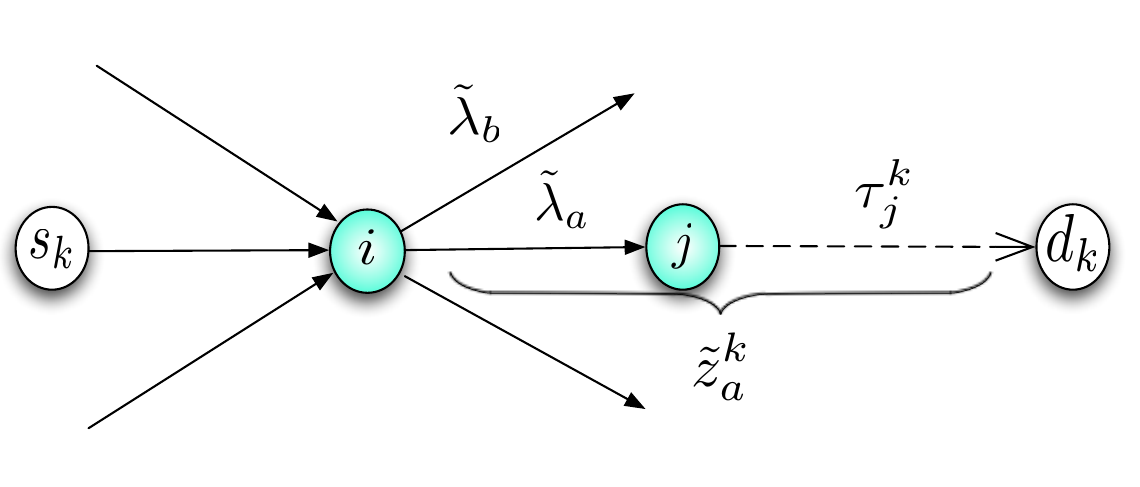}
\caption{\label{dp}Variables for dynamic programming equations}
\end{figure} 
If a packet at node $i$ is routed through the link $a\in A_i^+$ 
we have $\tilde \tau_i^k=\tilde\la_a+\tilde\tau_{j_a}^k$, so that a shortest path routing 
should choose the link with smallest  $\tilde\la_a+\tilde\tau_{j_a}^k$. 
Unfortunately, while the link delays $\tilde\la_a$ for $a\in A_i^+$ might be observed 
at node $i$, this is not the case for the $\tilde\tau_{j_a}^k$'s which depend on future 
delays that will be experienced when traversing the downstream links.
Suppose instead that  only the expected values $\tau_{j_a}^k$ are known and available at node $i$ and 
that each packet from source $k\in K$ observes the 
$\tilde\la_a$'s and is routed through the link $a\in A_i^+$ that minimizes  
$\tilde z_a^k=\tilde \la_a+ \tau_{j_a}^k$ to the next node $j_a$ where the process repeats.
Thus, denoting $E_a^k\triangleq\{\tilde z_a^k\leq \tilde z_b^k\,\,\, \forall b\in A_i^+\}$,
the packets from source $k\in K$ move across the network according to 
a Markov chain with transition probabilities
\begin{equation}
\label{matriz_trans}
P_{ij}^k=\left\{
\begin{array}{cl}
\PP(E_a^k) & \mbox{if } ij=a\\[1ex]
0									   & \mbox{otherwise} 
\end{array} \right.
\end{equation}
for $i\neq d_k$, while the destination $d_k$ is an absorbing state. 
The expected flows correspond to the invariant measures of these
Markov chains,  leading to a flow distribution rule in which 
the throughput flow $y_i^k$ from source $k$ that enters node $i$,
splits among the links $a\in A_i^+$ according to (see Figure \ref{dp2})
\begin{equation}\label{split}
v_a^k=y_i^k\; \PP(E_a^k).
\end{equation}
The throughputs $y^{k}=(y_i^k)_{i\neq d_k}$
can be computed from the stationary equations 
$y^k=\sum_{j=0}^{\infty} [(\hat P^{k})^{\p}]^j \delta^kx^k $,
where $\hat P^k=(P_{ij}^k)_{i,j\neq d_k}$ is the reduced transition matrix on the non-absorbing 
states, and $\delta_i^k=1$ for $i=s_k$ and $\delta_i^k=0$ otherwise. This may also be written 
as $y^k=x^k \delta^k+(\hat P^{k})^{\p} y^k$ which corresponds to 
the standard flow conservation equations
\begin{equation}
{ \label{cons_flujo1}
y_i^k=  x^k\delta_i^k + \mbox{$\sum_{a \in A_i^-}$} v_a^k}.
\end{equation}

These equations can be restated compactly using expected utility theory. Namely,
let us write $\tilde z_a^k=z_a^k+\epsilon_a^k$ as the sum of its expected value
$z_a^k= \la_a+ \tau_{j_a}^k$ plus a noise $\epsilon_a^k$ with $\EE(\epsilon_a^k)=0$, 
and assume that the distribution of $\epsilon_a^k$ does not change with $z_a^k$ (for a discussion 
of this assumption see \S\ref{future_work}).  Also, for simplicity 
$\epsilon_a^k$ is supposed to have continuous distribution so that the expected utility functions introduced 
next will be differentiable
(distributions with point masses can be treated as in \cite{Cominetti:2008}). Under these assumptions, the transition probabilities in \eqref{matriz_trans} can be expressed as
$\PP(E_a^k)=\frac{\partial \varphi_i^k}{\partial z_a^k} (z^k)$ where $ \varphi_i^k$ denote
the expected utility functions

\begin{equation}\label{euf}
\varphi_i^k(z^k)=\left\{\begin{array}{cl}
\EE ( \min\limits_{a\in A_i^+} \{z_a^k+\eps_a^k\} )&\mbox{if $i\neq d_k$}\\[2ex]
0&\mbox{if $i=d_k$}
\end{array}\right.
\end{equation}
which allow us to rewrite the flow equations \eqref{split}-\eqref{cons_flujo1} as
\begin{equation}
{\label{cons_flujo2}
\left\{
\begin{array}{ll}
v_a^k= y_i^k \, \frac{\partial \varphi_i^k}{\partial z_a^k} (z^k) &\,\, \forall a\in A_i^+ \\[1ex]
y_i^k = x^k \delta_i^k +  \sum_{a \in A_i^-} v_a^k &\,\, \forall i \neq d_k. 
\end{array}
\right.}
\end{equation}

On the other hand,  assuming that the cost-to-go variables $\{\tilde\tau_{j_a}^k:a\in A_i^+\}$ are 
independent from the local queueing times $\{\tilde\la_a:a\in A_i^+\}$, we may compute the expected 
value of $\tilde\tau_i^k$ by  conditioning on the events $E_a^k$ as
\begin{eqnarray*}
\tau_i^k=\EE(\tilde \tau_i^k)&=&\mbox{$\sum_{a\in A_i^+} \EE(\tilde \la_a+ \tilde\tau_{j_a}^k| E_a^k)\PP(E_a^k)$}\\
&=& \mbox{$\sum_{a\in A_i^+} \EE(\tilde \la_a+ \tau_{j_a}^k| E_a^k)\PP(E_a^k)$}\\
&=&\mbox{$\EE(\min\limits_{a\in A_i^+}\{\tilde \la_a+\tau_{j_a}^k\})$}
\end{eqnarray*}
so that
\begin{equation}
\label{DP0}
\left\{
\begin{array}{ll}
\tau_i^k= \vphi_i^k(z^k) & \forall i \in N\\[1ex]
z_a^k= \la_a + \tau_{j_a}^k  & \forall a\in A.
\end{array}
\right. 
\end{equation}

\vspace{2ex}
Under mild conditions it was proved in  \cite{Cominetti:2008} 
that, given the $\la_a$'s,  system \eqref{cons_flujo2}-\eqref{DP0}
has a unique solution $(v,y,\tau,z)$.
It was also shown that these equations, together with the equilibrium 
conditions $\la_a=s_a(w_a)$ where $w_a=\sum_{k\in K}v_a^k$ represents the total expected link load, have a unique solution 
$(\la,w,v,y,\tau,z)$ called a {\em Markovian Traffic Equilibrium} (MTE). This equilibrium 
is characterized by a pair of dual optimization problems analog to {\sc (p-w)} and {\sc (d-w)}.
As a matter of fact, the dual problem has exactly the same form 
$$
\min\limits_{\lambda\in\RR^A} \sum\limits_{a\in A} \int_{\la_a^0}^{\la_a} \!\!\!s_a^{-1}(z)\, dz-\sum_{k\in K}x^k \tau_k(\la)\leqno{\mbox{\sc (d-mte)}}
$$
where $\tau_k(\la)\triangleq\tau_{s_k}^k(\la)$ with $\tau_i^k(\la)$ the solution of \eqref{DP0}.

The expected utility maps $\varphi_i^k(\cdot)$ convey all the information required to describe a 
Markovian routing and may be considered as the primary modeling objects. These maps
are determined by the random variables $\epsilon_a^k$ which are ultimately tied to the arc
random costs $\tilde \la_a$.  The class $\mathcal{E}$ of maps 
that can be expressed in the form \eqref{euf}  admits an analytic characterization (see \cite{Cominetti:2008}): they are the $\mathcal{C}^1$
maps $\vphi:\RR^n\to \RR$ that are concave, componentwise non-decreasing, and which satisfy in addition
\begin{enumerate}
\item[(a)] $\vphi(x_1+c,\ldots,x_n+c)=\vphi(x_1,\ldots,x_n)+c$
\item[(b)] $\vphi(x)\to x_i$ when $x_j\to \infty$ for all $j\neq i$
\item[(c)] for $x_i$ fixed, $\frac{\partial \vphi}{\partial x_i}
(x_1,\ldots,x_n)$ is a continuous distribution function on the remaining variables.
\end{enumerate}
Note also that $\vphi(x)\leq\min\{x_1,\ldots,x_n\}$. 
In what follows we assume that the model is specified directly in terms of a family of maps 
$\varphi_i^k\in\mathcal{E}\!$ with $\varphi_{d_k}^k\equiv 0$. However, we note that these maps 
are not used explicitly by our distributed protocol in Section \S\ref{distributed-algorithm}.

\vspace{2ex}
\noindent{\sc Remark.} Since packet movements are governed by a Markov chain, cycling may occur 
and additional conditions are required to ensure that packets reach the destination with probability one.
A simple case is when source $k$ considers only the arcs in $A_i^+$ that lead closer to destination $d_k$
({\em e.g.} $\tau_{j_a}^k<\tau_{i_a}^k$), 
so that the corresponding Markov chain is supported over an acyclic graph $(N,A^k)$. To deal with this 
case it suffices to redefine
$$\varphi_i^k(z^k)\triangleq\EE(\min_{a\in A_i^{k+}}\{z_a^k+\epsilon_a^k\})$$
so that $P_{ij}^k=\mbox{$\frac{\partial\varphi_i^k}{\partial z_a^k}(z^k)$}=0$ for all 
$a\!=\!ij\!\not\in\! A_i^{k+}\!\!$.

\section{Rate control with Markovian routing}
\label{cross-layer}

We proceed to develop a cross-layer model that combines a NUM approach for rate control based on 
queueing delays, with a Markovian multipath routing based on total delays. 
Each source 
$k\in K$ is characterized by an origin $s_k$, a destination $d_k$, and a continuous decreasing rate 
function $f_k:(0,\infty)\to(0,\infty)$ with $f_k(q^k)\to 0$ as $q^k\to\infty$, while every link $a\in A$ has a continuous increasing 
latency function $s_a:[0,c_a)\to[\la_a^0,\infty)$ with $\la_a^0=s_a(0)\geq 0$. 
Packets are routed according to a Markovian strategy characterized by a family of 
maps $\vphi_i^k \in \EP$ with $\vphi_{d_k}^k\equiv 0$. Sources adjust
their rates as a function $x^k=f_k(q^k)$ of the total queueing delay 
$q^k= \tau_k(\la)-\tau_k^0$, where $\tau_k(\la)$ is the end-to-end expected
delay defined in the previous section and
 $\tau_k^0$ is the minimal travel time considering propagation delays only.

Informally, the source rates $x^k$ induce flows $v_a^k$ 
and total link loads $w_a$. These loads determine link expected delays $\lambda_a=s_a(w_a)$ that
yield end-to-end delays $\tau_k(\la)$ for each source and corresponding 
queueing delays $q^k$. At equilibrium, these queueing delays must induce the original 
rates $x^k=f_k(q^k)$.

\begin{defi}
A pair $(w,x)$ with $w=(w_a)_{a\in A}$ and $x=(x^k)_{k\in K}$ is called a 
\emph{Markovian Network Utility Maximization (MNUM) equilibrium}  if and only if
$w_a=\sum_{k \in K} v_a^k$\, where $(v^k,y^k)$ solve the flow conservation 
constraints  \eqref{cons_flujo2} with $(\tau^k,z^k)$ satisfying \eqref{DP0}, 
together with the link delay relations $\la_a=s_a(w_a)$ and the rate equilibrium 
conditions $x^k=f_k(q^k)$ where $q^k= \tau_k(\la)-\tau_k^0$.
\end{defi}

\subsection{Reduced formulation of MNUM}
\label{reduced-model} 
In order to establish the existence and uniqueness of equilibria we begin by reducing MNUM  
to an equivalent set of equations that involves only the variables $\la$. To this end we need to extend the 
results in \cite{Cominetti:2008} for which we consider a fixed non-negative link delay vector $(\la_a)_{a\in A}$. 
We first show that \eqref{DP0} uniquely defines $z^k$ and $\tau^k$ as implicit functions of $\la$. 
This system can be equivalently stated solely in terms of the variables 
$\tau^k$ as
\begin{equation}
{\label{DP2}
\tau_i^k= \vphi_i^k \left((\la_a+\tau_{j_a}^k)_{a\in A} \right)}
\end{equation}
so it suffices to prove that the latter uniquely defines $\tau^k$ as a function of $\la$. 

\begin{prop}
\label{Prop:conv_iter}
Let $k\in K$  and denote ${\bar \tau_i^k}$ the cost of a shortest path from $i$ to destination $d_k$ with link costs $\la_a$.
Suppose also that $\hat \tau^k\in\RR^N$ is such that
\begin{equation}
\label{S}\hat \tau^k_i\leq \vphi_i^k((\la_a+\hat \tau^k_{j_a})_{a\in A})\quad\quad(\forall\,i\in N).
\end{equation}
Then $\hat \tau^k\!\leq\! \bar \tau^k$ and moreover, starting from $\tau^{k,0}\!=\!\bar \tau^k$, the 
iterates computed by  
\begin{equation} 
{\label{iter_pf} \tau_i^{k,n+1}=\varphi_i^k((\la_a+\tau_{j_a}^{k,n})_{a\in A})}\quad\quad n=0,1,2\ldots
\end{equation}
are non-increasing and converge to a solution $\tau^k$ of \eqref{DP2} with $\tau_i^k\in [\hat \tau_i^k,\bar \tau_i^k]$.
\end{prop}

\begin{proof}
In order to prove that $\hat \tau^k\leq \bar \tau^k$ let ${\cal I}= \{i: \hat \tau_i^k\leq \bar \tau_i^k\}$ and 
suppose  by contradiction ${\cal I}\neq N$.
Since $\hat \tau_{d_k}^k\leq 0 = \bar \tau_{d_k}^k$ we have $d_k\in {\cal I}$. 
Consider a shortest path from a node $i_0\not\in{\cal I}$ to $d_k$, and let $i\not\in{\cal I}$ be the last node 
before entering ${\cal I}$ and $j\in {\cal I}$ the next node. Since $\vphi_i^k(z^k)\leq\min_{a\in A_i^+}z_a^k$ we get
\begin{eqnarray*}
\hat \tau_i^k &\leq& \varphi_i^k((\la_a+\hat \tau_{j_a}^k)_{a\in A})
\\& \leq&\la_{ij}+\hat \tau_j^k \leq \la_{ij}+ \bar \tau_j^k= \bar \tau_i^k < \hat \tau_i^k.
\end{eqnarray*}
This contradiction proves that $\hat \tau^k\leq\bar \tau^k$.

Let us prove next the convergence of the iteration \eqref{iter_pf}. We note that 
\begin{eqnarray*}
\tau_i^{k,1}&=&\vphi^k_i((\la_a+\tau_{j_a}^{k,0})_{a\in A})\\
&\leq& \min_{a\in A_i^+} \{\la_a+\bar \tau_{j_a}\} 
= \bar \tau_i = \tau_i^{k,0}
\end{eqnarray*}
from which it follows inductively  that the sequence \eqref{iter_pf} is non-increasing: if
$\tau^{k,n}\leq \tau^{k,n-1}$ then
\begin{eqnarray*}
\tau_i^{k,n+1} &=& \vphi^k_i((\la_a+\tau_{j_a}^{k,n})_{a\in A}) \\
&\leq& \vphi^k_i((\la_a+\tau_{j_a}^{k,n-1})_{a\in A}) = \tau_i^{k,n}. 
\end{eqnarray*}
It remains to show that the sequence $\tau^{k,n}$ is bounded below by $\hat \tau^k$.
We prove this by induction: the base case $\tau^{k,0}=\bar \tau^k\geq\hat \tau^k$ was just proved above,
while for the induction step it suffices to note that $\tau^{k,n} \geq\hat \tau^k$ implies
\begin{eqnarray*}
\tau_i^{k,n+1}&=& \vphi^k_i((\la_a+ \tau_{j_a}^{k,n})_{a\in A})\\
&\geq& \vphi^k_i((\la_a+\hat \tau_{j_a}^k)_{a\in A})\geq  \hat \tau^k_i.
\end{eqnarray*}
By continuity it follows that the limit of $\tau^{k,n}$ satisfies \eqref{DP2} and 
$\tau_i^k\in [\hat \tau_i^k,\bar \tau_i^k]$.
\hfill$\Box$\end{proof}

\vspace{2ex}
\noindent{\sc Remark.} The previous result gives a procedure to solve \eqref{DP2}:
compute the shortest path delays $\bar \tau^k$ and then iterate \eqref{iter_pf}.  
Alternatively one may start from $\tau^{k,0}=\hat \tau^k$ in which case the iterates increase and are bounded from above by $\bar \tau^k$, hence these iterates also converge to a solution of  \eqref{DP2}.

\vspace{2ex}
\begin{definition}
We denote $\mathcal{P}$ the set of all $\lambda \in \RR^{A}$ such that for each destination
$k\in K$ there exists $\hat \tau^k\in\RR^N$ satisfying
\begin{equation}\label{P}  
\hat \tau_i^k< \vphi_i^k((\la_a+\hat \tau_{j_a}^k)_{a\in A})\mbox{ for all }  i \neq d_k.
\end{equation}
\end{definition}

\vspace{1ex}
Note that $\mathcal{P}$ is an {\em open} convex domain, and for each 
$\la\in{\cal P}$ we have $\la'\in{\cal P}$ for all $\la'\geq\la$. 
In the sequel we extend the results in \cite{Cominetti:2008} which were based on
a much more stringent condition assuming that \eqref{P} holds with $\hat\tau^k=0$. 
The proofs differ substantially so we present them below.

\begin{lema} \label{lema:invertibilidad}
Let $\la\in{\cal P}$ and suppose that $(\tau^k,z^k)$ solves \eqref{DP0}. 
Let $\hat Q^k(z^k)$ be the matrix with entries $\hat Q_{ia}^k(z^k)={\partial\vphi_i^k\over\partial z_a^k}(z^k)$
for $i\neq d_k$ and $a\in A$.
Then
\begin{enumerate}
\item[(a)] For each $i\!\neq\! d_k$ there exists $j\!\in\! N$ such that 
$P_{ij}^k\!>\!0$ and $\hat\tau_j^k-\tau_j^k> \hat\tau_i^k-\tau_i^k$.
\item[(b)] The matrix $[I-\hat P^k(z^k)]$ is invertible.
\item[(c)] Equation \eqref{cons_flujo2} has a unique 
solution given by $v^k=\hat Q^k(z^k)^{'}y^k\!\geq\! 0$ with 
$y^k\!=\![I\!-\!\hat P^k(z^k)^{'}]^{-1} \delta^kx^k\!\geq\! 0$.
\end{enumerate}
\end{lema}
\begin{proof} 
(a) Since the $\vphi_i^k$'s are concave and differentiable we have
$$\begin{array}{rcl}
\hat \tau_i^k&<&\vphi_i^k((\la_a+\hat \tau_{j_a}^k)_{a\in A})\\
&\leq& \vphi_i^k(z^k) + \sum\limits_{a\in A} \frac{ \partial \vphi_i^k}{ \partial z_a^k}(z^k)(\hat \tau_{j_a}^k-\tau_{j_a}^k) \\
				  & =  & \tau_i^k + \sum\limits_{ij\in A_i^+} P_{ij}^k (\hat \tau_j^k- \tau_j^k)
\end{array}
$$
from which (a) follows directly.

\vspace{1ex}\noindent
(b) Given $i\neq d_k$ and using (a) inductively we can find
a finite sequence of nodes $i_0,\ldots,i_m$ with $i_0=i, i_m=d_k$ 
and $P_{i_k i_{k+1}}>0$. Thus, starting from $i$, the chain 
has a positive probability of reaching the absorbing state $d_k$ 
in a finite number of steps. This implies that $\hat P^k(z^k)^m$ 
is strictly submarkovian for $m$ large enough, and therefore 
$[I-\hat P^k(z^k)]$ is invertible.

\vspace{1ex}\noindent
(c) The first equation of \eqref{cons_flujo2} can be rewritten as $v^k=\hat Q^k(z^k)^{'}y^k$, which 
substituted into the second equation yields $y^k=\delta^kx^k+\hat P^k(z^k)^{'} y^k$, 
so that (b) implies $y^k = [I-\hat {P^k}(z^k)^{'}]^{-1} \delta^k x^k$. The non-negativity of these quantities
follows from the fact that $x^k \geq 0$ while the matrices $\hat Q^k(z^k)$ and $\hat {P^k}(z^k)$ have non-negative entries
with $[I-\hat {P^k}(z^k)^{'}]^{-1}=\sum_{m=0}^\infty[\hat {P^k}(z^k)^{'}]^m$. 
\hfill$\Box$\end{proof}

\vspace{3ex}The next result is the key to reduce the
MNUM equations to a system in the variables $\la$.

\begin{prop}
If $\la \in {\cal P}$ then, for each source $k\in K$, the system 
\eqref{DP0} has a unique solution $z^k=z^k(\la)>0$ and 
$\tau^k=\tau^k(\la)>0$. Moreover, the functions $\la\to \tau_i^k(\la)$ 
and $\la\to z_a^k(\la)$ are concave, smooth and component-wise 
non-decreasing.
\end{prop}

\begin{proof}
It suffices to show that \eqref{DP2} defines implicit maps $\la\mapsto \tau_i^k(\la)$ 
over the domain ${\cal P}$ which are well defined, concave, smooth, and monotone.
We already proved the existence of a solution  with $\tau_i^k\geq\hat \tau_i^k$. 
Let us prove its uniqueness.

\vspace{0.5ex}
\noindent \underline{\emph{Uniqueness}}: 
Let $k\in K$ and consider two solutions $(\tau^1,z^1)$ and $(\tau^2,z^2)$ for  \eqref{DP0}. 
Let $\alpha=\max_{i\in N} (\tau_i^2-\tau_i^1)$ and denote $N^{\ast}$ the set of nodes 
where the maximum is attained.  For every $i\in N^{\ast}$, the concavity of $\vphi_i^k(\cdot)$ gives
\begin{eqnarray*}
\tau_i^2=\vphi_i^k(z^2) &\leq& \vphi_i^k(z^1) +  \mbox{$\sum_{a\in A_i^+}\frac{\partial \vphi_i^k}{\partial z_a^k}$} (z^1) (z_a^2-z_a^1)\\
&=& \tau_i^1 +\mbox{$ \sum_{a\in A_i^+} \frac{\partial \vphi_i^k}{\partial z_a^k}$} (z^1) (\tau_{j_a}^2-\tau_{j_a}^1)
\end{eqnarray*}
and, since $(\tau_{j_a}^2-\tau_{j_a}^1)\leq \alpha$ while the partial derivatives add up to 1, we get
$$\alpha = \tau_i^2-\tau_i^1 \leq \mbox{$ \sum_{a\in A_i^+}\frac{\partial \vphi_i^k}{\partial z_a^k} (z^1)$}(\tau_{j_a}^2-\tau_{j_a}^1) \leq \alpha.$$
It follows that for every $a\in A_i^+$ such that 
$\frac{\partial \vphi_i^k}{\partial z_a^k} (z^1) >0$ 
we necessarily have $j_a \in N^{\ast}$. Combining this fact with Lemma \ref{lema:invertibilidad}(a),
we can find a finite sequence of nodes in $N^{\ast}$ starting at $i$ and ending at $d_k$. Hence 
$d_k\in N^{\ast}$  so that $\alpha=\tau_{d_k}^2-\tau_{d_k}^1=0$, which implies $\tau^2\leq \tau^1$. 
Exchanging the roles of $\tau^1$ and $\tau^2$ we get the converse inequality so that 
$\tau^1=\tau^2$ proving uniqueness.
			
\vspace{0.5ex}
\noindent \underline{\emph{Concavity}}:  For $\alpha\!\in\! (0,1)$ set $\la^{\alpha}=\alpha \la^1\!+(1\!-\!\alpha) \la^2$ 
and $\tau^{\alpha}\!=\!\alpha \tau^k(\la^1)\!+(1\!-\!\alpha) \tau^k(\la^2)$. Denote
$z^1_a=\la_a^1+\tau_{j_a}^k(\la^1)$,
$z^2_a=\la_a^2+\tau_{j_a}^k(\la^2)$, and $z^\alpha_a=\la_a^\alpha+\tau_{j_a}^{\alpha}$.
Then $z^\alpha=\alpha z^1+(1-\alpha)z^2$ and the concavity of $\vphi_i^k$ implies
$$\begin{array}{rcl}
\vphi_i^k(z^\alpha)  &\geq&  \alpha \vphi_i^k(z^1) + (1-\alpha) \vphi_i^k(z^2) \\
						&  =   & \alpha \tau_i^k(\la^1)+ (1-\alpha) \tau_i^k(\la^2) \\
	      				        &  =   & \tau_i^{\alpha}.
\end{array}
$$
This proves that $\hat \tau^k=\tau^{\alpha}$ satisfies condition \eqref{S} for $\la^\alpha$, and then 
Proposition \ref{Prop:conv_iter} gives $ \tau^k(\la^\alpha)\geq \tau^{\alpha}$
which yields precisely the concavity inequality
$$\tau^k_i(\la^{\alpha}) \geq \alpha \tau_i^k(\la^1)+ (1-\alpha) \tau_i^k(\la^2).$$

\vspace{0.5ex}
\noindent \underline{\emph{Smoothness}}: This is a direct consequence of 
the implicit function theorem. Indeed, noting that $\tau_{d_k}^k=0$
we may reduce \eqref{DP2} to a system in the variables 
$(\tau_i^k)_{i\neq d_k}$. The Jacobian of this reduced system 
is $[I-\hat P^k]$ which is invertible by Lemma 
\ref{lema:invertibilidad}(b), so the conclusion follows. 

\vspace{0.5ex}
\noindent \underline{\emph{Monotonicity}}:  Let $\la^1\in{\cal P}$ and take $\la^2\geq\la^1$ so that  $\la^2 \in {\cal P}$.
The monotonicity of $\vphi_i^k$ implies
\begin{eqnarray*}
\tau_i^k(\la^1) &=& \vphi_i^k((\la_a^1+\tau_{j_a}^k(\la^1))_{a\in A}) \\
&\leq& \vphi_i^k((\la_a^2+\tau_{j_a}^k(\la^1))_{a\in A})
\end{eqnarray*}
Hence $\hat \tau^k=\tau^k(\la^1)$ satisfies \eqref{S} for $\la^2$, and
Proposition \ref{Prop:conv_iter} implies $\tau^k(\la^2)\geq \tau^k(\la^1)$.
\hfill$\Box$\end{proof}

\vspace{2ex}
The implicit maps $\tau^k(\la)$ and  $z^k(\la)$ defined by \eqref{DP0},
allow us to restate the MNUM equations solely in terms of the link delay vector $\la$.
Indeed, let $q^k(\la)=\tau_{s_k}^k(\la)-\tau_k^0$ and define $\tilde x^k(\la)=f_k(q^k(\la))$.
According to Lemma  \ref{lema:invertibilidad}(c) the equations \eqref{cons_flujo2} have 
unique solutions $v^k=v^k(\la)$ and $y^k=y^k(\la)$. Denoting  
$\tilde w_a(\la)=\sum_{k\in K}v_a^k(\la)$, the MNUM equations are equivalent to the 
reduced system of equations
$$
\la_a=s_a(\tilde w_a(\la))\quad\forall a\in A.
\leqno {\mbox{ \sc (r-mnum)}}$$

\subsection{Variational characterization}
\label{variational-model} 

We show that the reduced system {\sc (r-mnum)} corresponds to the optimality conditions of an
optimization problem which is a combination of the variational characterizations 
{\sc (d-num)} and {\sc (d-mte)}.

\begin{teo} \label{teorema:form_var}
Assume that $\la^0\in {\cal P}$. Then $(x^{\ast},w^{\ast})$ is an MNUM  equilibium 
iff $x^*=\tilde x(\la^*)$ and $w^*=\tilde w(\la^*)$ with $\la^*$ an optimal solution of the strictly convex program 
$$\min\limits_{\la \in {\cal P}}\,\, \Phi(\lambda)\,\,\triangleq\,\,  \displaystyle \sum\limits_{a\in A} \int_{\la_a^0}^{\la_a}\!\!\! s_a^{-1}(z)\,dz - \sum\limits_{k\in K}F_k(q^k(\la))\leqno{\mbox{\sc (d-mnum)}}$$
where $F_k(\cdot)$ denotes a primitive of $f_k(\cdot)$.
\end{teo}

\begin{proof}
Since $q^k(\la)=\tau_{s_k}^k(\la)-\tau_k^0$ is concave and the $f_k$'s are positive and decreasing,
it follows that the map $\Phi_1(\la)\triangleq-\sum_{k\in K}F_k(q^k(\la))$ is convex.
Also, since the $s_a^{-1}$'s are increasing we obtain that $\Phi_0(\la)\triangleq\sum\limits_{a\in A} \int_{\la_a^0}^{\la_a}s_a^{-1}(z)\,dz$
is stricty convex, and then so is $\Phi(\la)=\Phi_0(\la)+\Phi_1(\la)$. Hence, since ${\cal P}$ is open and convex, an optimal solution for
 {\sc (d-mnum)} is characterized by $\nabla \Phi(\la)=0$. Now
\begin{eqnarray*}
\mbox{$\frac{\partial \Phi}{\partial \la_a}$} &=& \mbox{$s_a^{-1}(\la_a) - \sum\limits_{k\in K} f_k(q^k(\la)) \frac{\partial q^k}{\partial \la_a}(\la)$}\\
& = &\mbox{$s_a^{-1}(\la_a)  -   \sum\limits_{k\in K} \tilde x^k(\la) \frac{\partial \tau_{s_k}^k}{\partial \la_a}(\la).$}
\end{eqnarray*}
An implicit differentiation of  \eqref{DP2} gives
$\frac{\partial \tau^k}{\partial \la}=\hat Q(\la)+\hat P^k(\la)\frac{\partial \tau^k}{\partial \la}$ so that
$\frac{\partial \tau^k}{\partial \la_a}=[I-\hat P^k(\la)]^{-1}\hat Q(\la)\ind_{a}$,
from which we get 
$$ 
\begin{array}{rl}
\tilde x^k(\la) \frac{\partial \tau_{s_k}^k}{\partial \la_a}(\la)&=\tilde x^k(\la) \ind_{s_k}^{'} [I-\hat P^k(\la)]^{-1}\hat Q(\la)\ind_{a} \\
		              &= ( [I\!-\!\hat {P^k}(\la)^{'}]^{-1} \delta^k \tilde x^k(\la))^{'} \hat Q(\la)\ind_{a} \\
	      		       &= {y^k}(\la)^{'} \hat Q(\la) \ind_{a}\\
		              &= v^k(\la)^{'} \ind_{a} \\
				&= v_a^k(\la) .
\end{array}
$$
Therefore $\frac{\partial \Phi}{\partial \la_a} = s_a^{-1}(\la_a)  - \tilde w_a(\la)$ and the optimality condition $\nabla\Phi(\la)=0$
coincides with the {\sc (r-mnum)} characterization of equilibria.
\hfill$\Box$\end{proof}

\vspace{2ex}
This characterization allows us to prove the existence and uniqueness
of an MNUM equilibrium.
\begin{teo} 
\label{prop:existencia_var}
Problem {\sc (d-mnum)} is strictly convex and coercive, hence it has a unique optimal solution
and therefore there exists a unique MNUM equilibrium.
\end{teo}

\begin{proof}
We already showed that the objective function $\Phi(\lambda)$ is strictly convex. 
Moreover, since $q^k(\la)$ is componentwise non-decreasing while the term 
$\int_{\la_a^0}^{\la_a} s_a^{-1}(y)\,dy$ is decreasing for $\la_a<\la_a^0$, it follows that the minimum
of   {\sc (d-mnum)} belongs to the set $\{ \la\geq\la^0 \}$. 
Hence, in order to establish coercivity, it suffices to show that the recession function
satisfies $\Phi^\infty(\lambda)>0$ for all $\la\geq 0$ with $\lambda\neq 0$.
Now,
\begin{eqnarray*}
\Phi_0^{\infty} (\la)&=& \lim_{t\to \infty}{1\over t}\Phi_0(t\la)\\
&=&\sum_{a\in A}\lim_{t\to\infty} \frac{1}{t} \int_{\la_a^0}^{t \la_a} \!\!\!\!s_a^{-1}(z)\, dz \\
&=& \sum_{a\in A}\lim_{t\to \infty}  s_a^{-1}(t\la_a)\la_a\\
&=&\sum_{a\in A}c_a\la_a>0. 
\end{eqnarray*}

\noindent In order to compute $\Phi_1^\infty(\lambda)$ let us consider the convex maps
$U_k(r)=-F_k(-r)$ and $h_k(\la)=- q^k(\la)$. 
It can be shown (see \cite{Cominetti:2008}) that  $h_k^{\infty}(\la)=-\bar \tau_{s_k}^k$
with $\bar \tau_{s_k}^k$ the cost of a shortest path with link delays $\la_a$.
Since $U_k^{\infty}(r)=0$ for $r\leq0$, we get $(U_k\circ h_k)^{\infty}(\la) = U_k^\infty(h_k^\infty(\la))=0$ for all $\la\geq 0$,
and then $\Phi_1^\infty(\la)=0$. 
As a consequence, for all directions $\la\geq 0$ with $\la\neq 0$ we get $\Phi^\infty(\la)>0$, so that 
 {\sc (d-mnum)} is inf-compact and therefore it has an optimal solution. 
\hfill$\Box$\end{proof}


\section{A distributed algorithm for MNUM}
\label{distributed-algorithm}

This section briefly describes how the MNUM framework can lead to
a distributed protocol for rate congestion control under Markovian multipath routing.
This protocol can be interpreted as a distributed algorithm 
that solves the variational problem {\sc (d-mnum}). 
The algorithm is based on a Markovian routing process for packets, with 
a slow update of the end-to-end expected delays $\tau_i^k$'s. This process is combined 
with a fast TCP adaptation of user's rates by estimating the end-to-end queueing delays 
to reach the equilibrium rates $\tilde x^k(\la)$. A more detailed description and analysis 
of the distributed protocol will be the subject of a forthcoming paper \cite{Guzman:2012}. 

\subsection{Packet routing based on local queues} \label{subsec:routepricing}

We adapt the ideas of \S\ref{Markov_routing} in order to define a routing
policy based on local information. To do this, for each packet with destination $k$
router $i$ must find the outgoing link $a\in A_i^+$ that realizes the minimum of
the values $\tilde z_a^k=\tilde\la_a+\tau_{j_a}^k$,  by adding the propagation delay 
$\la_a^0$, plus the current queueing delay $\tilde p_a$  of the link,  plus the estimate
of the expected delay $\tau_{j_a}^k$ advertised in the routing table of the next hop $j_a$. 
If destination $k$ is not reachable through $j_a$  we take $\tau_{j_a}^k$ as infinity (or a 
very large value).

This requires that each router periodically advertises its routing table providing an estimate of the total expected 
delays $\tau_i^k$ to all its known destinations. These estimates may be updated on a slow time-scale by  averaging 
the observed delays $\min_{a\in A_i^+} \{\tilde\la_a+\tau_{j_a}^k\}$ over a fixed number of  packets
or over all packets sent over a fixed time window. The observed average $\check\tau_i^k$ is used 
to update the estimate of the expected delay as 
$$\tau_i^k\leftarrow (1-\alpha)\tau_i^k+\alpha\check\tau_i^k.$$

\subsection{TCP protocol} \label{subsec:TCPprotocol}

 The TCP protocol for source rate control requires a mechanism by which every source $k$ can 
estimate $q^k(\la)=\tau_{s_k}^k(\lambda)-\tau_k^0$. 
Here we rely on the two time-scales assumption: sources control their rates at a much faster 
pace than routers, so they see link delays as constant. For fixed expected delays $\la$, the 
expected forward time coincides with $\tau_{s_k}^k(\la)$ so that, using standard protocols for 
estimating the forward time, sources can get an unbiased estimation $T_t^k$ of this total expected delay.

We also need a mechanism to estimate $\tau_k^0$, that is, the shortest distance considering propagation delay only and 
no queuing. Since the physical transmission speeds are roughly constant, these values are 
more stable and will only change when the network topology is modified by addition or removal of a router or link. 
One option is to estimate $\tau_k^0$ by the minimum forward time observed 
over all sent packets, just as in the single-path implementation of Vegas. The hope is that at 
least one packet will be routed along a shortest path and find no queues along its way. 
However, this estimator is known to be biased if the network is 
already congested and therefore it provides only an upper bound for $\tau_k^0$.  An alternative 
is to let each packet accumulate the propagation delay along its route and feedback the total to the 
source in the corresponding {\em ack} so that $\tau_k^0$ can be estimated as the minimum
value observed. As a third option one could implement a RIP protocol, in parallel with the estimation 
of $\tau_i^k$ in \S  \ref{subsec:routepricing}, to compute shortest paths using propagation delay as metric. 
This requires feedback from routers to sources and provides distance estimates for aggregate destinations 
(autonomous systems) which differ from the end-to-end minimum times $\tau_k^0$ by a small constant 
access time. This, however, does not invalidate the analysis in \S \ref{cross-layer}. 
We believe this option is an accurate and efficiently implementable one.

An alternative approach, proposed in  \cite{Paganini:2009}, is to replace the {\em minimum} $\tau_k^0$
and use instead the {\em average} end-to-end propagation delay as the reference {\em baseRTT} 
in the Vegas protocol. This average propagation delay can be estimated directly by sources using 
the moving average technique described in \cite{Paganini:2009} or, alternatively, by adapting the
randomized ECN protocol in \cite{Adler:2002} for estimating an additive measure
for single-path routing. A generalization of this method to multipath routing is developed 
in \cite{Guzman:2012}. While both approaches avoid the computational overhead of a 
RIP protocol, we must note that the average propagation delays  depend on the routes
being currently used which are themselves affected by the network congestion so that this base RTT
measure is not  flow-independent  as required in the analysis of section \S \ref{cross-layer}.

{\color{blue}

}

Finally, the free-flow times $T_k^0$ together with the unbiased estimators $T_t^k$ of 
$\tau_{s_k}^k$ for every packet $t$ arriving to destination, can be used to adjust the rates
by a stochastic approximation algorithm of the form
\begin{equation} 
\label{grad_estoc} x_{t+1}^k \leftarrow (1-\delta)x_t^k + \delta f_k(Q_t^k)
\end{equation}
where $Q_t^k=T_t^k-T_k^0$.
If we let sources adapt long enough so that $Q^k_t\sim q^k$ and $x_t^k\sim f_k(q^k)$, 
we can then proceed to update the router estimates of the end-to-end delays $\tau_i^k$.

\subsection{Complexity and implementation considerations}

The proposed schemes require little modifications to current TCP/IP protocols. The total expected
delay update for routers described in \S \ref{subsec:routepricing} has similar computational,
communication and memory requirements as distance-vector protocols such as RIP. 
The main difference is that the hop-count metric is replaced by the expected delay for 
each destination known to the router. The routing tables and expected delays are updated in 
a moderated time-scale and advertised as usual to neighboring routers. 
For other implementation considerations see 
\cite{Paganini:2009}, where the authors make a full description on how to generalize the hop-count
distance by other congestion prices.

Let us stress that our proposed method for estimating the propagation delay requires a parallel
computation by RIP-type protocols, which conveys a memory overhead on routers to store 
an additional metric for each entry in the routing table. This might be costly but permits an 
accurate estimation of queuing delays. 
The simpler alternative of accumulating propagation delays on the packet headers has 
also a storage overhead, although it can be efficiently implemented 
in some situations \cite{Adler:2002}. Which of these approaches has the best cost-benefit tradeoff 
will be the subject of another study  \cite{Guzman:2012}.

Concerning TCP, the estimation of forward travel times and the ECN estimation are standard 
features of TCP/IP protocols. A relevant issue is that routing along paths with heterogeneous 
delays may induce packet reordering so that the duplicate ACK feature of TCP must be turned off,
allowing for some buffering at the receiver to reorder the packets (see \cite{Paganini:2009} for details). 
However, since our routing strategy is based on minimizing end-to-end delay, in steady state
packets should arrive approximately in order so that excessive buffering should not be needed. 
An alternative to avoid packet reordering is to keep individual TCP-connections single path and use 
a hashing technique to distribute these connections for load balancing  (see {\em e.g.} 
\cite{cao:2000,Paganini:2010}). The downside of this technique is its coarser granularity which 
may cause instabilities in the load balancing and routing.

\section{Comparison with related work}
\label{related-work}

Multipath routing can be decomposed into two main tasks: computation of paths and 
traffic splitting. The splitting can be controlled directly by sources or in a decentralized 
manner by routers. Moreover, it can be implemented either on a per-packet basis 
(each packet following a possibly different path) or a per-flow basis where each 
TCP-connection is assigned a single path and load balancing is achieved by distributing 
these connections among paths. Accordingly, several alternative approaches have been proposed 
in the literature. We briefly compare MNUM with some of the previous works. For more 
complete surveys and discussions of the challenges involved in multipath routing 
we refer to \cite[Gojmerac]{Gojmerac:2007}, \cite[He and Rexford]{He:2008} and
\cite[Lee and Choi]{Lee:2002}.

The seminal paper \cite[Gallager]{Gallager:1977} introduced a distributed 
routing protocol that finds an optimal multi-commodity flow minimizing average delays. The model 
considers flow dependent latencies, but traffic demand and network topology are assumed to 
be fixed. In a similar context, the PEFT protocol in \cite[Xu {\em et al.}]{Xu:2011} develops a routing 
scheme based on exponential penalties using link-prices specifically tuned to reproduce an optimal 
multi-commodity flow. Although PEFT operation is distributed, flow optimization and link-price 
tuning require centralized computation. For large networks where the topology and the traffic 
change continuously during operation, this involves substantial processing and communication
 overheads. In contrast, MNUM  automatically 
adjusts the routing to variations in traffic and topology in a decentralized manner. 
Incidentally, we note that  the MNUM routing also 
takes the form of an exponential penalty if link costs are distributed Gumbel, although MNUM deals 
directly with the actual randomness of the links without assuming any specific a priori distribution. 

Multipath routing with elastic traffic was considered in \cite[Kelly {\em et al.}]{Kelly:1998}. 
In this setting each source directly controls the flow rates to be sent over a set of available paths, 
using a TCP-like feedback mechanism based on congestion signals. The framework uses
fluid-flow dynamics that converge to an optimal solution of a multipath 
version of NUM, which provide a template for designing packet-level protocols. 
A variant of this model including feedback delays is the basis for the {\em overlay TCP} scheme 
proposed in \cite[Han {\em et al.}]{Han:2003}. The effect of delays is also studied in 
\cite[Kelly and Voice]{Kelly:2005} providing sufficient conditions for dynamical stability,
while \cite[Key {\em et al.}]{Key:2011} analyzes the asymptotic behavior as the number of 
connections increase. 
With a similar goal, \cite[Lin and Shroff]{Lin:2006} consider a discrete iteration for solving multipath 
NUM using a variant of the  proximal point algorithm that yields a decentralized 
algorithm with good convergence properties. The implementation of source-controlled 
routing schemes presupposes that the network supports multipath routing. One alternative for this is 
to use Label Switched Path tunnels using MPLS \cite[Villamizar]{Villamizar:1999}, however the processing 
overhead of per-flow routing does not scale well with the number of connections. A different option is 
considered in overlay TCP by establishing a set of overlay routers at some peering 
points, with traffic between these points controlled by standard single-path protocols. While this favors incremental 
deployment, path diversity is limited to the extent that traffic must be routed through the predetermined peers. 
In our approach any neighboring router can play the role of an overlay node, and no per-flow 
routing is required since traffic splitting is controlled by routers rather than sources. It is also worth noting 
that in contrast with the approaches that start from NUM and fluid-flow dynamics which then lead to a 
packet level protocol, we proceed in the reverse order from packet dynamics to its equilibrium described
by MNUM.

Traffic splitting controlled by routers was considered in \cite[Paganini]{Paganini:2006} and
\cite[Paganini and Mallada]{Paganini:2009} by combining a routing scheme as 
in \cite[Gallager]{Gallager:1977} with a rate control as in \cite[Kelly {\em et al.}]{Kelly:1998}. 
Flow splitting is decentralized at each router $i$ by using split ratios $(\alpha_a^k)_{a\in A_i^+}$ that control 
the fraction of packets from source $k$ that are forwarded along each outgoing link $a\in A_i^+$. 
These ratios  are dynamically adjusted so that the routing concentrates over the links that 
belong to currently shortest paths. The framework uses a fluid-flow model from which a packet-level 
protocol is derived. Our approach is similar in the sense that flow splitting is also decided locally at routers 
based on current delays, so that congestion aware paths are selected automatically. 
In both approaches all the paths are potentially available and only the currently optimal ones are 
used, although other path choice rules  can also be incorporated by forbidding flow on some links
(see the remark at the end of \S\ref{Markov_routing}). A difference between both approaches 
is that while
\cite{Paganini:2006,Paganini:2009} is based on expected values of queuing delay, our routing 
evolves stochastically using the current state of local queues and considering total delay 
including queueing plus propagation. One reason for this choice is to allow a finer per-packet 
granularity in load balancing, keeping packet reordering under control. 
In contrast, since \cite{Paganini:2006,Paganini:2009} uses paths with heterogenous total 
delays, load balancing is implemented at a coarser per-flow granularity by using hashing \cite{Paganini:2010}.
As a final remark, the protocol in  
\cite{Paganini:2009} requires three time-scales to ensure convergence: a fast source rate adaptation, 
a medium time-scale for route price updates, and a slow update of splitting ratios. 
We only use two time-scales: a slow one for estimating the delays and a fast one 
for rate control and routing.

\section{Conclusions and future work}
\label{future_work}

We proposed a new cross-layering model for TCP/IP control under multipath routing. 
The motivation for our routing mechanism comes from using local information about 
queueing delays as well as the expected delays from the next hops to the destination, in order to 
exploit the available capacity by sending packets through several alternative routes. 
To achieve this purpose, we considered a Markovian routing 
combined with a Vegas-like TCP protocol for rate control. The routing process was characterized by 
studying the expected dynamic programming equations which lead to a Markovian Traffic Equilibrium,  
together with a standard Network Utility Maximization model for the TCP steady state. This led to a 
variational characterization of the equilibrium that allowed us to prove its existence and uniqueness,
and which inspired a distributed protocol for attaining the equilibrium. 

There are several unsolved issues. Firstly, further 
research is required to provide a theoretical support for the convergence of these protocols.
A detailed analysis should study the relation between the 
packet-level dynamics and our flow-level model. Our equilibrium model relies on this assumption: namely, 
we base our updates on aggregated flow information
as well as in the two time-scales convergence of equilibrium flows. 
A complete analysis should explain to which extent the flow 
model captures the packet level dynamics, and how fast the equilibrium
flows are attained by sources. Interesting recent results along this line can be found 
in \cite{Kelly:2009,Walton:2009}.

Another interesting question is related to the model of randomness assumed.
We considered an additive structure $\tilde z_a^k=z_a^k+\epsilon^k_a$ which
presumes the same variability of delays regardless of the average flow levels observed. 
A more realistic model should consider higher variability for higher expected delays,
based either on a detailed analysis of the distribution of waiting times at queues, or at least 
using a simplified multiplicative randomness model of the form $\tilde z_a^k=z_a^k(1+\epsilon_a^k)$. 

A final line of research has to do with simulating this protocol
in a realistic environment. A fair comparison with single-path routing requires the
presence of uncertainty and delays in  information transmission.
Simulation may provide an idea on the effective increase in performance
that one might expect from a Markovian multipath routing.


\bibliographystyle{spmpsci}      
\bibliography{Bibliography}   

\end{document}